# Tuning of Quantum Paraelectricity of M-type Hexaferrite $BaFe_{12}O_{19}$ by External Parameters

Jing Zhang[1]*, Feng Peng[2]*, Na Su[3], Long Zhang[1], Yugang Zhang[1], Young Sun[3], Rujun Tang[2‡], Yisheng Chai[1,3†]

[1]*Low Temperature Physics Laboratory, College of Physics, Chongqing University, Chongqing 401331, China*

[2]*Jiangsu Key Laboratory of Thin Films, School of Physical Science and Technology, Soochow University, Suzhou 215006, People's Republic of China*

[3]*Center of Quantum Materials and Devices, Chongqing University, Chongqing 401331, China*

†yschai@cqu.edu.cn

‡tangrj@suda.edu.cn

* J. Z. and F. P. contributed equally to this work

## Abstract

M-type hexaferrite $BaFe_{12}O_{19}$ was recently reported to be a new type of quantum paraelectrics with triangular lattice by showing a low temperature dielectric plateau due to quantum fluctuation. It has also been proposed to have a possible quantum-dipole-liquid ground state. To suppress its quantum fluctuations and reach a possible quantum critical point, we have tuned its quantum paraelectricity in three ways: (i) $^{57}$Fe isotope replacement; (ii) in-plane compressive strain; and (iii) hydrostatic pressure. It is found that 95% $^{57}$Fe replacement and the in-plane strain are more effective to drive its ground state closer to a critical region by inducing a peak feature in the temperature dependence of dielectric constant. In contrast, the application of hydrostatic pressure pushed the system away from the quantum critical point by gradually suppressing the plateau feature in dielectric constant. Our combined efforts reveal the potential of the M-type hexaferrites for studying the quantum critical behaviors.

# I. INTRODUCTION

Quantum paraelectricity has been an interesting topic for decades. A well-known case is $SrTiO_3$ (STO), which displays a very large dielectric constant at low temperatures but fails to go through a ferroelectric transition due to quantum fluctuations associated with the zero-point energy [1]. The temperature dependent relative dielectric constant $\varepsilon_r$ can be depicted by the Barrett formula which is derived from a transverse field Ising model of pseudospin: [2].

$$\varepsilon_r = A + \frac{M}{\frac{1}{2}T_1 coth\left(\frac{T_1}{2T}\right) - T_0} \tag{1}$$

where A is a constant, $T_0$ reflects the effective coupling constant between electric dipole moments and the positive or negative sign represents ferro- or antiferroelectric interactions, respectively. $T_1$ (=$\Omega/k_B$, $\Omega$ is transverse field) is the quantum tunneling possibility. $M = nk_B\mu^2$, where $n$ is the density of dipoles and $\mu$ denotes the local dipolar moment. If the quantum fluctuation in STO can be suppressed, its ground state will pass a quantum critical point (QCP) and be tuned to an ordered ferroelectric phase. Various stimulus such as electric field, uniaxial stress, and isotopic substitution can fulfil such a task [3-6].

Recently, some of M, Z, and W-type hexaferrites are identified as a new type of magnetic quantum paraelectric materials with $Fe^{3+}$ ions induced local electric dipoles [7,8]. Hexaferrites are iron oxides with hexagonal structures. Depending on their chemical formula and crystal structures, hexaferrites can be classified into six main types [9,10]: M-type (Ba, Sr)$Fe_{12}O_{19}$, Y-type (Ba, Sr)$_2Me_2Fe_{12}O_{22}$, W-type (Ba, Sr)$Me_2Fe_{16}O_{27}$, X-type (Ba, Sr)$_2Me_2Fe_{28}O_{46}$, Z-type (Ba, Sr)$_3Me_2Fe_{24}O_{41}$, and U-type (Ba,Sr)$_4Me_2Fe_{36}O_{60}$, where Me is a bivalent metal ion. The M-type hexaferrite has the simplest structure to host the quantum paraelectricity, e.g., $BaFe_{12}O_{19}$ with $P6_3/mmc$ space group [11], as shown in Fig. 1(a). It can be described by a periodically stacking sequence of two basic building blocks ($S$ and $R$ blocks) with an overlap of hexagonally and cubically packing along the $c$-axis, two $S$ and two $R$ blocks are 180° rotations around the $c$ axis respectively, giving the unit cell structure $SRS^*R^*$ (* represents 180° rotating around *c* axis). The $Fe^{3+}$ ions forms tetrahedra, octahedral and bipyramidal by combining with $O^{2-}$. The layer containing $Ba^{2+}$ is a mirror plane.

The electric dipoles in $BaFe_{12}O_{19}$ are caused by the local displacement of $Fe^{3+}$ along the *c*-axis in the bipyramidal site. Therefore, only the *c*-direction relative dielectric constant $\varepsilon_r$ of single crystal samples shows quantum paraelectric behavior [7]. Later, Y. Sun *et al.* found that the $FeO_5$ bipyramidal unit in $BaFe_{12}O_{19}$ forms a perfect two-dimensional triangular lattice, pointing to a more exotic quantum-dipole-liquid state due to the existence of geometric frustration and quantum fluctuation [8]. In its magnetic counterpart of the antiferromagnet transverse field Ising model in a triangular lattice, theoretical calculations predict a phase diagram far from that of STO [12-14]. If the quantum fluctuation can be weakened, the magnetic system will go through a QCP in the universality class of three-dimensional XY model and enter the ordered ground state of spin dimer, and the symmetry of the system will be broken. At finite temperatures, there is a Kosterlitz–Thouless (KT) phase transition between the disordered phase and the bond-ordered phase. Therefore, it is highly desirable to find suitable approach to suppress the quantum fluctuation/quantum tunneling probability in $BaFe_{12}O_{19}$ to probe the possible QCP and other exotic states as a dipole version of antiferromagnet transverse field Ising model in a triangular lattice.

From the perspective of quantum mechanics, the quantum tunneling probability $T_1$ of $Fe^{3+}$ propagating through the bipyramidal equatorial plane (energy barrier) is expressed as follows:

$$T_1 = De^{-\frac{2a}{\hbar}\sqrt{2m(U_0-E)}} \tag{2}$$

where $D$ is a constant, $a$ is the barrier width, $m$ is the mass of $Fe^{3+}$, $U_0$ is the barrier height, and $E$ is the particle energy. According to Eq. 2, there are several strategies to reduce the $T_1$. First, one can enhance the mass. When the isotope $^{57}Fe$ is used to fully replace $^{56}Fe$, $T_1$ will be reduced by about 0.88%. Second, one can increase the barrier width or height by the external pressure to reduce $T_1$. The above proposed approaches can be tested to tune the ground state of $BaFe_{12}O_{19}$ closer to or pass the expected QCP.

In this work, we have tried to tune the quantum tunneling probability and subsequently quantum paraelectricity behaviors of $BaFe_{12}O_{19}$ by $^{57}Fe$ isotope replacement, in-plane substrate mismatch and hydrostatic pressure. It is found that only nearly 100 percent $^{57}Fe$ isotope replacement and in-plane compressive stress can drive the ground state of this system into a critical region. The application of hydrostatic pressure will push the system away from the critical behavior and significantly suppress the electrical dipole moment.

## II. MATERIALS AND METHODS

The $Ba(^{56}Fe_{1-x}{}^{57}Fe_x)_{12}O_{19}$ ($x$=0-0.95) polycrystalline samples were synthesized by solid-state reaction. $BaCO_3$ (purity 99%), $^{56}Fe_2O_3$ (purity 99.9%), $^{57}Fe_2O_3$ (purity 95.47%) precursors were weighted in the 1:(6−6$x$):6$x$ nominal mole ratios. Firstly, moderate alcohol is added in the precursor, mixed for 10 hours by mixing machine, then grinded in an agate mortar for 30 minutes. The fine powder was sintered and grinded at 1140 °C for 4 hours. Finally, these powders were pelletized into thin cylindrical platelets and sintered again at 1200 °C for 96 hours in the air. The epitaxial $Ba(^{56}Fe_{1-x}{}^{57}Fe_x)_{12}O_{19}$ ($x$=0 and 0.95) films with 200 nm thick were deposited on the conductive Nb-doped (111)-oriented $SrTiO_3$ (lattice constant $a$ = 3.905 Å) substrates by pulsed laser deposition using $Ba(^{56}Fe_{0.05}{}^{57}Fe_{0.95})_{12}O_{19}$ platelets synthesized above [15]. The quality of the polycrystal and thin film samples were checked using room temperature X-ray diffraction (XRD, PANalytical X'Pert Powder, Cu Kα, λ = 1.5406 Å) and transmission electron microscope (TEM, Tecnai FEI G 2 F20). Silver paste was deposited on the two larger polished surfaces of platelet acting as electrodes for dielectric measurements. From thin film samples, gold spots with a diameter of 300 μm are deposited onto the film as top electrodes while the Nb-doped STO as the bottom electrodes for the dielectric measurements. Relative dielectric constant $\varepsilon_r$ was measured under frequencies ($f$) of 100 kHz and 1 MHz using an LCR meter (Agilent E4980A). The temperature ($T$) was controlled using a Dynacool system (Quantum Design). Dielectric properties under hydrostatic pressure were performed in a BeCu piston pressure cell using Daphne 7373 as the transmitting medium.

## III. RESULTS AND DISCUSSION

To check the quality of the $BaFe_{12}O_{19}$ specimen, the room temperature XRD of thin films and bulk samples are measured, as shown in Figs. 1(b) and 1(c), respectively. All the peaks can be indexed to M-type hexaferrite and no impurity phase is found in any samples. The XRD reciprocal space mapping (RSM) in Fig. 1(d) shows narrow diffraction pattern of (008) peak. The TEM image in Fig. 1(e) shows sharp interface between the $BaFe_{12}O_{19}$ film and STO substrate. The dotted selected-area diffraction pattern further proves that the film is well epitaxially grown on STO. The in-plane lattice mismatch between $BaFe_{12}O_{19}$ and STO substrate is about 6% [15]. The lattice constants of all the samples are calculated accordingly, as shown in Table I. There is no significant change on the lattice constants by $^{57}Fe$ isotropic doping in bulk samples. However, the *c*-axis of the film samples have been elongated clearly, indicating that the in-plane stress by STO substrate effectively reduces the in-plane lattice constant and stretches the *c*-axis simultaneously.

To investigate the effects of isotope doping on the quantum paraelectricity of $BaFe_{12}O_{19}$, the *T*-dependent relative dielectric constant $\varepsilon_r$ of bulk $Ba(^{56}Fe_{1-x}{}^{57}Fe_x)_{12}O_{19}$ is measured. For comparison, the normalized relative dielectric constant $\varepsilon_r/\varepsilon_{\max}$ curves are shown in Fig. 2 at two frequencies (100 kHz and 1 MHz). For $x$=0, 0.33 and 0.66 samples, $\varepsilon_r$ increases with the decrease of temperature, and exhibits a plateau below 6 K, which is consistent with the quantum paraelectric behavior of *c*-axis dielectric constant in single-crystal $BaFe_{12}O_{19}$ [7]. There is no frequency dependence in the temperature range studied for $x\leq0.66$ samples, implying good insulating nature of these bulk samples. For $x = 0.95$ sample, the quantum paraelectricity still exists down to 2 K. However, $\varepsilon_r$ is slightly frequency dependent in this sample, indicating a more conducting nature for this doping level. When we expand the low temperature region below 8 K for all the samples (the insets of Fig. 2), a gradual change from the plateau to a peak feature by $^{57}Fe$ doping can be found. This peak feature may indicate a precursor of a QCP.

To further reveal the effects of $^{57}Fe$ doping on the quantum paraelectric behaviors, $\varepsilon_r/\varepsilon_{max}$ curves at $f$= 1 MHz between 2 K and 80 K are fitted with the Barrett formula in Eq. (1) and the obtained parameters are summarized in Table II. There is no systematic difference of $T_0$, $T_1$ and $M$ values among $x$ = 0, 0.33 and 0.66 bulk samples, while in $x$ = 0.95, their magnitudes in $T_0$ and $T_1$ decrease significantly, and the $M$ value at least doubles. The decrease in the magnitude of $T_0$ reflects a weaker dipole-dipole interaction between heavier $Fe^{3+}$ ions. The doubling of $M$ in $x$=0.95 indicates a larger off-center displacement of $^{57}Fe$ than that of $^{56}Fe$. Together with the decrease of 6% in $T_1$, they are consistent with our expectation that heavier $^{57}Fe$ ions lead to a smaller quantum tunneling possibility. The enhancement of $M$ must be intrinsic since a conducting sample will lead to a suppression of the dielectric constant in cooling, like the case in less insulating $SrFe_{12}O_{19}$ single crystal [7]. Therefore, the high level of $^{57}Fe$ doping seems to push the ground state of quantum paraelectricity closer to the expected QCP.

Since the nearly full isotope replacement of $^{57}Fe$ is not enough to push the ground state to the QCP, we now turn to the external pressure as the tuning parameters. There are two possible ways to apply pressures which can tune the dipole moment/energy barrier and change the tunneling probability—the first is in-plane pressure ($P_{\text{in-plane}}$) and the second is hydrostatic pressure ($P_{\text{Hydro}}$), as shown in Fig. 3a. For $P_{\text{in-plane}}$, with the shrink of $a$-axis, the Fe ions will be pushed away further from the center, resulting in a larger dipole moment/larger energy barrier and smaller tunneling probability. In our epitaxially grown films with $x$ = 0 and 0.95, it is expected to employ strong $P_{\text{in-plane}}$ to the films due to lattice mismatch. For the $P_{\text{Hydro}}$, both $a$ and $c$ axes will be suppressed while its consequence is unclear. In previous studies, hydrostatic pressure on $SrTiO_3$ can push the ground state away from its QCP [16].

Then, to demonstrate the effects of $P_{\text{in-plane}}$, the $T$-dependent $\varepsilon_r/\varepsilon_{max}$ of film samples $x$ = 0 and $x$=0.95 are measured and shown in Figs. 3(b) and 3(c), respectively. Both films show quantum paraelectric behaviours with the clear peak feature around 6 K. This peak feature is similar to that of the $x$=0.95 bulk sample and indicates a similar ground state close to the QCP. In particular, there is negligible frequency dependence in the temperature range studied for the $x$=0 film sample. Both curves are fitted with Eq. (1) and the obtained parameters are shown in Table II. For film samples, both $T_0$ and $T_1$ are larger than that in bulk samples, which may be due to the largely oriented grains in the films. The enhancement of $M$ is even more significant (2.5-8.8 times larger) in film samples than that in bulk counterparts. This difference cannot be explained by the random and oriented grains in poly and film samples, respectively, but by the $P_{\text{in-plane}}$ from the STO substrate. Note that, the larger $M$ value in $x = 0$ film suggest a larger $P_{\text{in-plane}}$ which cannot be well controlled during film growth process while the effect of $^{57}$Fe isotope is not obvious in this case.

To test the effect of hydrostatic pressure, we measured the $T$-dependent $\varepsilon_r/\varepsilon_{\max}$ at $f = 1000$ kHz of $x = 0.95$ poly sample under selected $P_{\text{Hydro}}$, as shown in Fig. 3(d). The hydrostatic pressures of 0, 0.6, 1.8, 3.8, 7.8, 7.9, 17.6 and 29.2 kbar are applied. The impact of hydrostatic pressure is manifested in two aspects. Firstly, in the low temperature range, the hydrostatic pressure strongly suppresses the magnitude of $\varepsilon_r/\varepsilon_{max}$ which indicates a continuously reduced dipole moment in the bipyramidal $Fe^{3+}$ ions, as shown in the right panel of Fig. 3(a). Secondly, the low temperature peak feature gradually changes to the plateau feature by the application of hydrostatic pressure. Under the maximum 29.2 kbar, the plateau can persist up to 15 K. These behaviors show that hydrostatic pressure plays the same role in regulating the quantum critical behavior in $SrTiO_3$ [16] so as to push the ground state away from the possible QCP.

Finally, we turn our attention to the phase diagram of $Ba(^{56}Fe_{1-x}{}^{57}Fe_x)_{12}O_{19}$ under external tuning parameters. As the quantum paraelectricity can be described by a transverse Ising model, it is possible that the M-type hexaferrite with antiferroelectric interaction and triangular lattice can borrow the phase diagram predicted by the transverse field Ising spin model on triangle lattice, as shown in Fig. 4. According to the previous theoretical studies [12-14], this model has a bond-ordered ground state in low transverse field and is superimposed by a KT phase at finite temperature. With increasing the transverse field, the system undergoes a quantum phase transition that its QCP is in the three dimensional-XY universality class. In larger transverse fields, the quantum paraelectric phase comes in. Above the QCP, there is a broad temperature region where the quantum critical behaviors dominate. Based on this phase diagram, the position of each sample can be inferred, as shown in Fig. 4. The polycrystal $BaFe_{12}O_{19}$ is located at the right side of QCP and outside of the critical region. Through the isotope replacement and in-plane pressure, $x = 0$, $x = 0.33$ and $x = 0.66$ are in the quantum paraelectric phase outside of the critical region while $x= 0.95$ and thin film samples are within the critical region with a peak feature in dielectric constant. By applying the hydrostatic pressure, the $x= 0.95$ poly sample is moved away from the critical region to the normal quantum paraelectric phase.

## IV. CONCLUSION

We have applied three kinds of approaches on the $BaFe_{12}O_{19}$ to tune its quantum paraelectricity, namely: i) $^{57}Fe$ isotope replacement; ii) in-plane compressive stress from the STO substrate; iii) hydrostatic pressure. It is found that only nearly 100 percent $^{57}Fe$ isotope replacement and in-plane compressive stress can tune the ground state of this system into a critical region. The application of hydrostatic pressure will push the system away from the critical behavior and significantly suppress the electrical dipole moment. In order to reach the QCP in this system, further efforts are required.

## ACKNOWLEDGMENTS

This work is supported by the Natural Science Foundation of China under grant Nos. 11674384, 51772200, 51725104, and the National Key Research and Development Program of China (Grant No. 2021YFA1400303). We Would like to thank Miss G. W. Wang at Analytical and Testing Center of Chongqing University for her assistance.

# FIGURE CAPTIONS

**Figure 1 |** (a) The crystal structure of $BaFe_{12}O_{19}$. Room temperature XRD 2θ patterns of (b) film samples and (c) bulk samples of $Ba(^{56}Fe_{1-x}{}^{57}Fe_{x})_{12}O_{19}$ ($x$=0,0.3,0.66 and 0.95). (d) XRD reciprocal space mapping (RSM) pattern of $Ba(^{56}Fe)_{12}O_{19}$ thin film around $BaFe_{12}O_{19}$ (008) peak. (e) TEM cross-sectional image and of $Ba(^{56}Fe)_{12}O_{19}$ thin film with selected-area diffraction pattern inserted.

**Figure 2 |** The temperature dependent normalized relative dielectric constant $\varepsilon_r/\varepsilon_{max}$ of (a) $x$=0, (b) $x$=0.33 (c) $x$=0.66 and (d) $x$=0.95 for polycrystal bulk, and (f) $x$=0 and (e) $x$=0.95 for film samples $Ba(^{56}Fe_{1-x}{}^{57}Fe_{x})_{12}O_{19}$ under two frequencies.

**Figure 3 |** (a) Schematic pressure configurations in lattice. The temperature dependent normalized relative dielectric constant $\varepsilon_r/\varepsilon_{max}$ of (b) $x$=0 and (c) $x$=0.95 for $Ba(^{56}Fe_{1-x}{}^{57}Fe_{x})_{12}O_{19}$ film with 100 kHz and 1000 kHz. (d) The temperature dependent capacitance under different hydrostatic pressure $P$ with $f$ = 1000 kHz for bulk sample $x$ = 0.95.

**Figure 4 |** The phase diagram of $BaFe_{12}O_{19}$ by varying the quantum tuning parameters.

# FIGURES

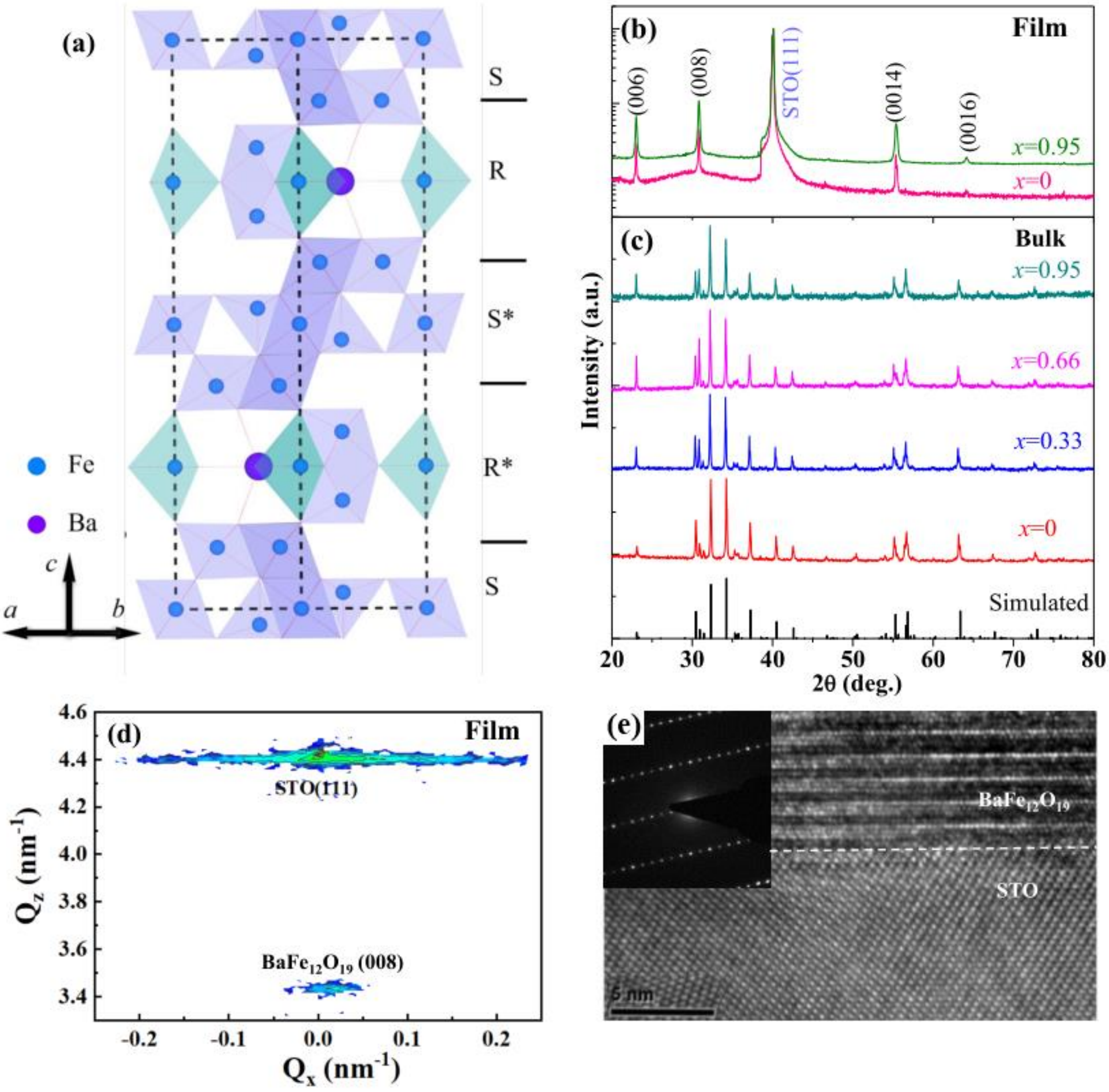

Figure 1 Jing Zhang *et al*.

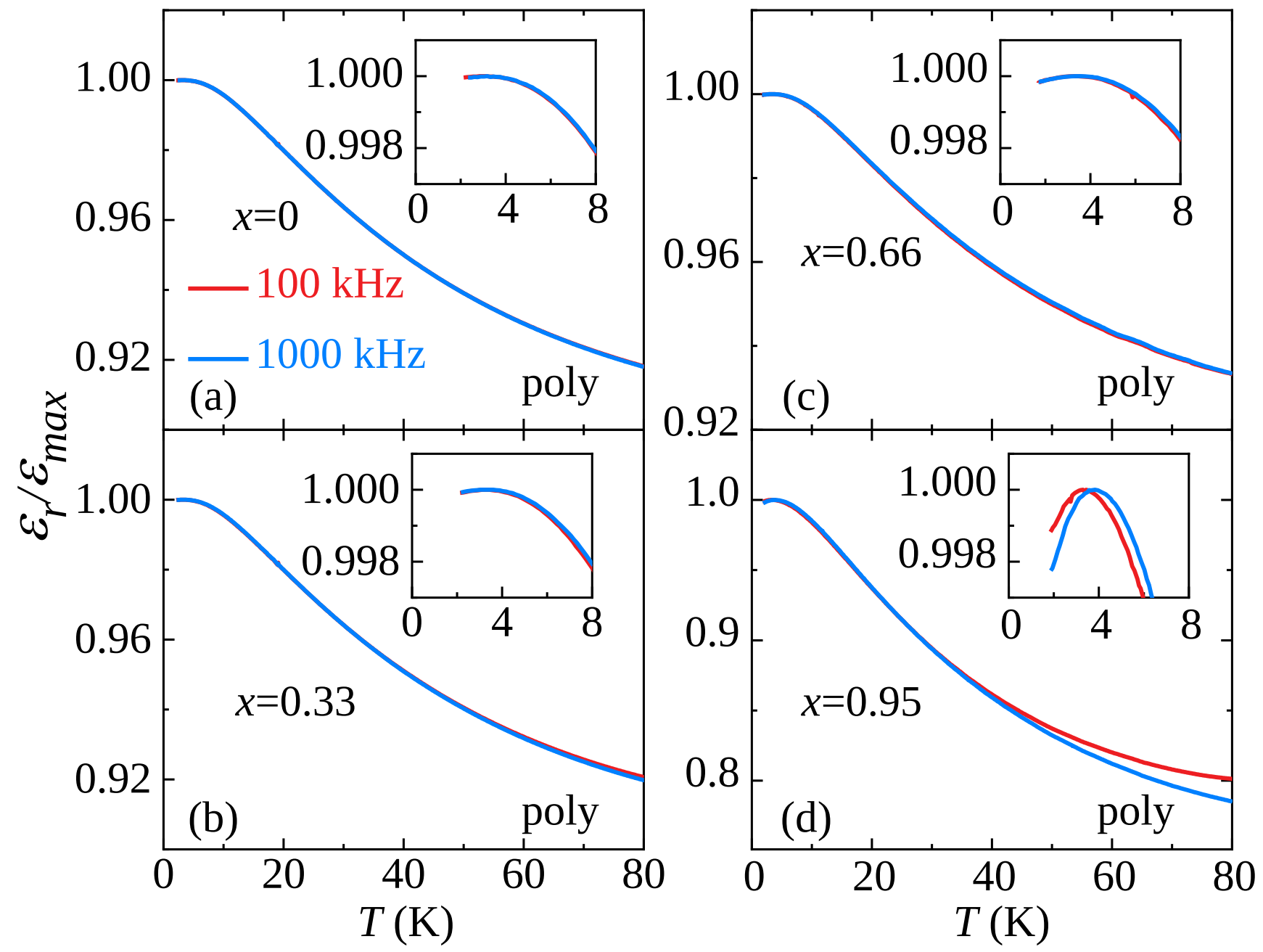

Figure 2 Jing Zhang *et al.*

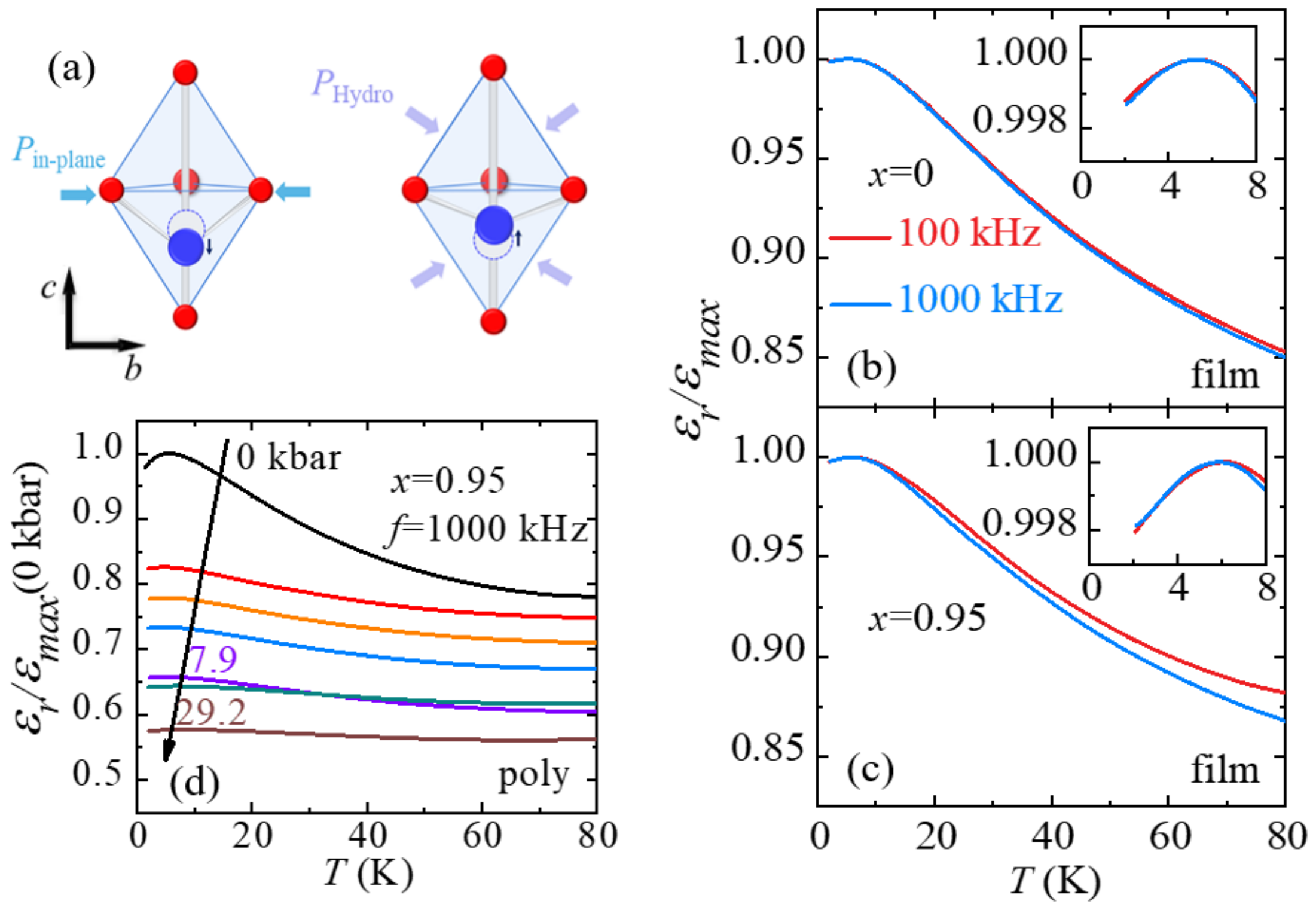

Figure 3 Jing Zhang *et al*.

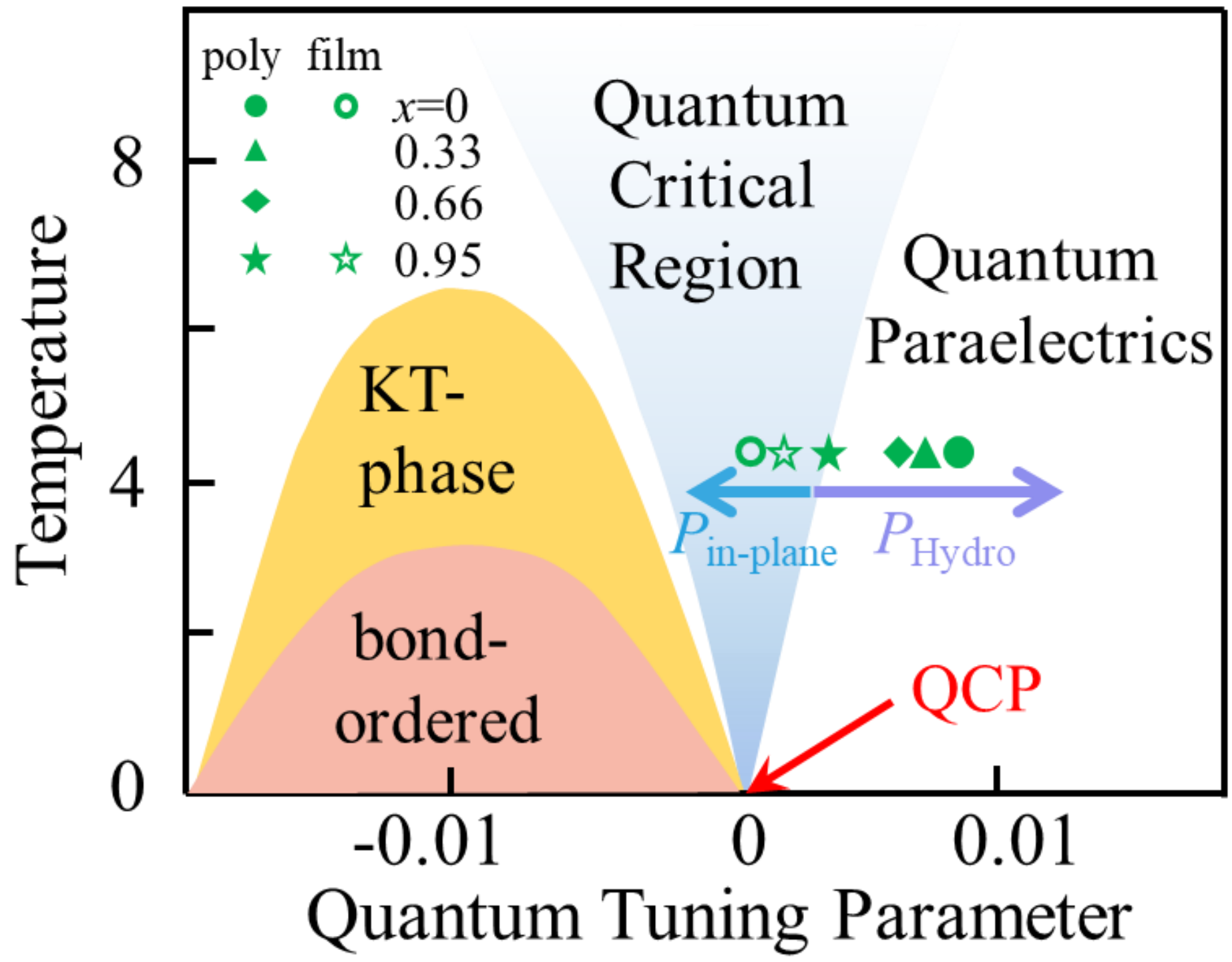

Figure 4 Jing Zhang *et al*.

# TABLES

Table I. The calculated lattice constants *a* and *c* of $Ba(^{56}Fe_{1-x}{}^{57}Fe_x)_{12}O_{19}$.

| $Ba(^{56}Fe_{1-x}{}^{57}Fe_x)_{12}O_{19}$ | $x$=0 (poly) | $x$=0.3 (poly) | $x$=0.66 (poly) | $x$=0.95 (poly) | $x$=0 (film) | $x$=0.95 (film) |
|---|---|---|---|---|---|---|
| *a* (Å) | 5.880 | 5.912 | 5.901 | 5.903 | | |
| *c* (Å) | 23.170 | 23.196 | 23.204 | 23.192 | 23.275 | 23.244 |

Table II. The fitting parameters $T_0$, $T_1$, and $M$ to Eq. (1) below 80 K of $Ba(^{56}Fe_{1-x}{}^{57}Fe_x)_{12}O_{19}$ with $f$=1000 kHz.

| $Ba(^{56}Fe_{1-x}{}^{57}Fe_x)_{12}O_{19}$ | $x$=0 (poly) | $x$=0.3 (poly) | $x$=0.66 (poly) | $x$=0.95 (poly) | $x$=0 (film) | $x$=0.95 (film) |
|---|---|---|---|---|---|---|
| $T_0$ (K) | -32.17 | -32.76 | -32.53 | -23.30 | -52.72 | -47.84 |
| $T_1$ (K) | 33.42 | 32.51 | 32.63 | 30.63 | 38.21 | 36.85 |
| $M$ (K) | 134.17 | 155.53 | 111.53 | 297.76 | 1183.75 | 790.12 |